\documentstyle[aaspp]{article}
\input psfig.tex
\lefthead{Spaans et al.}
\righthead{Search for Interstellar Water}

\begin{document}

\title{Search for Interstellar Water in the Translucent Molecular Cloud
toward HD 154368}

\author{Marco Spaans\footnote{Hubble Fellow, currently at the Harvard-Smithsonian Center for Astrophysics, 60 Garden Street, Cambridge, MA 02138}, David Neufeld}
\affil{Department of Physics \& Astronomy, The Johns Hopkins University, Baltimore, MD 21218}
\author{Stephen Lepp}
\affil{University of Nevada, 4505 S.\ Maryland Parkway, Las Vegas, 
NV 89154--4002}
\author{Gary J. Melnick, and John Stauffer}
\affil{Harvard-Smithsonian Center for Astrophysics, 60 Garden Street,
Cambridge, MA 02138}

\begin{abstract}

We report an upper limit of $9\times 10^{12}$ cm$^{-2}$
on the column density of water in the translucent cloud along 
the line-of-sight toward HD 154368.  This result is based
upon a search for the C-X band of water near 1240 \AA\
carried out using the Goddard High Resolution Spectrograph
of the Hubble Space Telescope.
Our observational limit on the water abundance together with detailed
chemical models of translucent clouds and previous measurements of OH
along the line-of-sight constrain the branching ratio in the
dissociative recombination of H$_3$O$^+$ to form water.
We find at the $3\sigma$ level that no more than 30\% of 
dissociative recombinations of H$_3$O$^+$ can lead to H$_2$O.  

The observed spectrum also yielded high-resolution observations
of the Mg II doublet at 1239.9 \AA\ and 1240.4 \AA, allowing
the velocity structure of the dominant ionization state
of magnesium to be studied along the line-of-sight.  The
Mg II spectrum is consistent with GHRS observations at lower
spectral resolution that were obtained previously but
allow an additional velocity component to be identified.

\end{abstract}


\section{Introduction}

Translucent clouds have total extinctions $A_{\rm V}$ in the range of 2-5
magnitudes.  The physical and chemical conditions that characterize
translucent clouds  are therefore intermediate between those in 
diffuse and in dense molecular clouds (Crutcher 1985; van Dishoeck \& 
Black 1988).  Photodissociation rates are significantly smaller in 
translucent clouds than in diffuse clouds, and the column densities 
of molecules like CO, OH and CS are correspondingly larger. 
Translucent clouds can be observed through 
absorption lines of CN, CH, CH$^+$ and C$_2$ as well as through 
millimeter emission lines of CO and CS.
Although they are an abundant constituent of dense molecular clouds 
(Jacq et al.\ 1988; Zmuidzinas et al.\ 1995;
Gensheimer et al.\ 1996; van~Dishoeck \& Helmich 1996;
Cernicharo et al.\ 1997) and are expected to be 
the dominant coolant of warm dense regions within such clouds
(Neufeld \& Kaufman 1993; Neufeld, Lepp \& Melnick 1995),
water molecules have not been detected in diffuse or
translucent molecular clouds.  In this paper, we report the results
of a search for water in the translucent cloud along the line of sight to 
HD 154368.

HD 154368 is an O9.5 Iab star at an estimated distance of 800 pc 
from the Sun (Snow et al.~1996)
that is situated
near the Sco OB 1 association (Blades
\& Bennewith 1973). The line-of-sight toward HD 154368 lies about $\rm 14^o$ from
the core of the $\rho$ Oph molecular cloud,
which is located at a distance of 125$\pm$25 pc (de Geus, de Zeeuw, \& Lub
1989)
and has a heliocentric radial velocity of $\rm -6.6\, km \,s^{-1}$.
The main H I (e.g.\ Riegel \& Crutcher 1972) and Na I (e.g.\ Crawford,
Barlow, \& Blades 1989) absorption features are observed at
a radial velocity that is close to that of the $\rho$ Oph molecular cloud, 
a result that suggests that the main component of gas toward HD 154368 is
probably not close to the star and is more likely to be the outer envelope
of a dense molecular cloud
only about 125 pc from the Sun. The color excess along
the line of sight is E(B--V)=0.82 and most of the gas toward HD 154368 resides
in two clouds centered near --3.26 (main component)
and --20.95 km s$^{-1}$ (heliocentric).

This line-of-sight has been observed extensively
from the ground by means of 
narrow H I 21 cm absorption features (e.g.~Riegel \& Crutcher 1972);
optical absorption lines of 
Na~I~D (e.g.~Crawford et al.~1989),
CN B--X (0,0) and A--X (0,0), CH, CH$^+$; and
near-infrared absorption lines of C$_2$ in the A--X Phillips system at 8750
\AA\ (van Dishoeck \& de Zeeuw 1984). The CH observations can be used to infer
the total column density of H$_2$ along the line-of-sight (Gredel,
van Dishoeck, \& Black 1993), a quantity that is needed to determine the atomic
depletions. The red system of CN has been used by Gredel, van Dishoeck,
\& Black (1991) to derive an electron density of
0.05-0.15 cm$^{-3}$ for the molecular component. Molecular emission lines
are also detectable toward HD 154368.
Data on $^{12}$CO $J=1-0$, $J=2-1$, and $J=3-2$ and $^{13}$CO $J=1-0$ have
been presented by
van Dishoeck et al.~(1991), and have been used to constrain the column density
of CO and the density of H$_2$ in the molecular component. These authors
confirmed the relatively low density $n_{\rm H}\approx 350$ cm$^{-3}$ derived
independently from the C$_2$ absorption data. The $^{12}$CO $J=1-0$ 
distribution over a region
of $30'\times 30'$ around the star is featureless, a result that
suggests once more
that the cloud and the star are not located close to one another.

The velocity structure of the gas toward HD 154368 is well known through the
Na I data obtained using the Ultra High Resolution Facility 
at the Anglo Australian Telescope (Snow et al.~1996) and high
resolution Ca II observations (Crawford 1992). The Na I data indicates seven
velocity components at --27.7, --20.95, --18.2, --14.75, --10.5, --3.26, 
and +5.62
km s$^{-1}$ where 96\% of the neutral sodium resides in the --3.26 km s$^{-1}$
feature. The Ca II results show five velocity components at --27.6, --20.9,
--14.4, --4.3, and +6.8 km s$^{-1}$. These five coincide roughly with the Na I
components, but their relative column densities are different. Approximately
50\% of the ionized calcium resides in the --4.3 km s$^{-1}$ feature with
an even distribution over the remaining velocity components.

A detailed discussion of all these observations and the implications they
have for the physical conditions of the ambient medium can be found in Snow et
al.~(1996). In this work, their results will be adopted and the main focus
will be on the resulting OH and H$_2$O chemistry.

\section{Observational Results}

Observations of HD 154368
in the 1234.49--1241.35 \AA\ spectral region were obtained 
at a spectral resolution of $\rm \sim 3\,km\,s^{-1}$ using 
Echelle-A of the Goddard High Resolution Spectrograph
(GHRS) on HST.  
The data, obtained with the use of standard substepping,
were reduced within the IRAF environment following procedures
developed by the GHRS Science Team (Cardelli, Ebbets, \& Savage 1990; 1993).
The equivalent widths and wavelengths of the features in the calibrated
spectra were obtained with various IRAF routines and in particular the
multi-component ``specfit'' package developed by Kriss (1994) that allows
for deblending.

Figures 1a and 1b show the region around the C\ $^1B_1$-X\ $^1A_1$ band of
water near 1240 \AA.  We find no evidence in the spectrum for
features at the expected location of
either the ortho lines $1_{01}-2_{11}$
(1239.382 \AA) and $1_{01}-1_{11}$ (1240.153 \AA), or the para line
$0_{00}-1_{10}$ (1239.728 \AA). 
To determine a reliable upper limit in the
presence of noise, an artificial water feature was added until a $3\sigma$
identification could be made with the specfit program.
The assumed line width for the artificial feature was 
$b=2.9 \rm\, km \,s^{-1}$,
the measured $b$-value for the dominant MgII component (see Table 1
below).
With the water oscillator strengths of
Smith \& Parkinson (1978) and Smith et al.~(1981), and the assumptions
that one-quarter and three-quarters of the water molecules are respectively
in the $0_{00}$ and $1_{01}$ states, we find a $3\sigma$ upper
limit of $9\times 10^{12}$ cm$^{-3}$ 
for the column density of water toward HD 154368. This upper limit was
obtained by co-adding the separate lines in the spectrum, and was not
found to change when the more recent
line rest wavelengths of Watson
(1997, private communication) and the oscillator strengths of Watson
(1997, private communication) and Yoshino (1997, private communication) were
used. 
The 3$\sigma$ upper limits on the equivalent widths of the 1239.382 \AA,
1239.728 \AA\ and 1240.153 are 0.85 m\AA, 0.89 m\AA\ and 0.84 m\AA,
respectively.  
For comparison, the limit on the H$_2$O column density in the more diffuse
($A_{\rm V}\approx 1$ mag) $\zeta$ Oph interstellar cloud is $3\times 10^{12}$
cm$^{-2}$ (Snow 1976).

Figure 2 shows the Mg II doublet features
around 1239.9 \AA\ and 1240.4 \AA.
Table 1 presents the results of the fitting
procedure for Mg II and lists the heliocentric
velocities of the various components,
their $b$ values, equivalent widths and column densities. 
The fit is based upon the line rest wavelengths of Morton (1991) and the
oscillator strengths of Fitzpatrick (1997).
Our results for the equivalent widths are in good general agreement with those
obtained previously from lower resolution spectra of Mg II
(Snow et al.~1996), although our derived column densities are a factor of 2.4
smaller due to the use of the Fitzpatrick (1997) oscillator strengths, rather
than those tabulated by Morton (1991).
A systematic velocity shift of $\sim 2$ km s$^{-1}$ is present. This shift
is not inconsistent with the GHRS wavelength calibration uncertainty and is
most likely caused by geo-magnetic effects. We also find an additional velocity
feature at --36.2 km s$^{-1}$, apparent in both components of the doublet,
with a column density $\sim 1 \times 10^{14}\rm cm^{-3}$, which 
has not been detected previously.

Figure 3 shows a clear 
detection of the Ge II transition at 1237.059
\AA. The heliocentric velocity of this component
is --1.2 km s$^{-1}$, in excellent
agreement with that of the main Mg II component. 
No other GeII features fall in the spectral region covered by 
our observations.

\section{Model}

To interpret our upper limit on the water column density
toward HD 154368, we used the numerical code 
described by Spaans (1996) and by Spaans \& van Dishoeck (1997) to
construct chemical models of the foreground cloud.  A
detailed description of the numerical model is given in these references and
the interested reader is referred there for details. The model input
consists of the density (assumed constant) $n_{\rm H}$, the total extinction
through the cloud $A_{\rm V}$, the gas phase elemental abundances, the
strength of the illuminating radiation field $I_{\rm UV}$ in units of the
average interstellar radiation field of the Galaxy, the shape of the
ultraviolet extinction curve, the cosmic ray ionization
rate $\zeta$, and the branching ratios for the dissociative
recombination
of H$_3$O$^+$. The cloud is modeled as a two-sided plane parallel slab.
Dust attenuation and self-shielding of H$_2$ and CO are included explicitly
through a Monte Carlo radiative transfer method. The extinction curve of
Snow et al.\ (1996) is adopted for the specific line of sight toward
HD 154368, and is extended to wavelengths of 1000 \AA\ using the method of
Cardelli, Clayton, \& Mathis (1989) with a value of the ratio of total to
selective extinction of $R_{\rm V}=3.1$.

The model produces as output the thermal and chemical structure of the cloud
and the column densities of all major chemical species. Figures 4a and 4b
present
the results for a model with $n_{\rm H}=325$ cm$^{-3}$, $I_{\rm UV}=3$,
$\zeta =5\times 10^{-17}$ s$^{-1}$, a H$_3^+$ recombination rate of
$8\times 10^{-7}T^{-0.7}$ cm$^3$ s$^{-1}$ (Smith \& Spa\~nel 1993) and
the depletions as given in Snow et al.~(1996),
which reproduces the observations to within 50\% for all observed species
except CH$^+$. The CH$^+$ problem
is well known in interstellar studies and no
attempt is made here to alleviate it (e.g.\ Dalgarno 1976; Black 1988).
The C$_2$ observations of van Dishoeck
\& de Zeeuw (1984) yield a kinetic gas temperature of 25 K in the molecular
region of the cloud ($A_{\rm V}>1$ mag). The temperature dependence depicted in
Figure 4b is consistent with this value. The dust temperature is assumed to be
constant and does not strongly influence the chemistry.

The predicted abundances of water and OH depend critically upon
two poorly-known parameters to which the abundances of other 
observed species are insensitive: the cosmic ray 
ionization rate, $\zeta$, and the branching
ratio for dissociative recombination of H$_3$O$^+$.  We define 
$f \rm (H_2O)$, $f\rm (OH)$, and $f\rm (O)$ as the fractions of dissociative
recombinations of H$_3$O$^+$ that lead to H$_2$O, OH and O.

With the cloud geometry, density and radiation field fixed 
by observations of other species,  
one can construct a grid of models with different
values of $\zeta$ and $f$(H$_2$O). 
The OH and H$_2$O column densities predicted in these simulations are
shown in Figure 5 (solid lines). For these models we have assumed that
$f$(O)=0 (or equivalently, $f {\rm (H_2O)} + f\rm (OH) = 1$), as has been
assumed in most theoretical studies and is motivated
by the recent experiment of Vejby-Christensen et al.~(1997).
The
cross 
indicates the 3 $\sigma$ limit on the water
column density reported here and the observed OH column density of
N(OH)=$1.4\times 10^{14}$ cm$^{-2}$ (Black 1997, private communication).
From these results it follows at the 3$\sigma$ confidence level
that $f$(H$_2$O) can be no larger than 0.30 and $\zeta$ no 
larger than $1 \times 10^{-16}\rm\, s^{-1}$. This rules out the large
$f$(H$_2$O) branching ratio $\sim 1$ predicted by 
Bates (1986; but note the revised estimates of Bates 1991),
provided that our adopted water chemistry, and 
in particular the value of the H$_3^+$ recombination rate, is secure. 

Two recent experiments have yielded contradictory results
for the H$_3$O$^+$ branching ratios.
The experiment of Vejby-Christensen et al.~(1997) 
implied that $f\rm (OH)=0.66$ and $f\rm(H_2O)=0.33$, values
that are formally ruled out by our limit on the
water column density, at least 
given the model assumptions discussed above.
The results of Williams et al.~(1996), by contrast, 
suggested that the dissociative recombination of 
H$_3$O$^+$ has a significant
channel to O ($f\rm (O)= 0.30$) and that the
channel to H$_2$O is much less important
($f \rm (H_2O) = 0.05$). To investigate the influence of the
channel to O, we have run additional models
in which $f {\rm (H_2O)} + f\rm (OH) = 0.7$
rather than unity.  The dashed curve in Figure 5 denotes the results for
these computations with $f \rm (H_2O) = 0.05$.  For $f\rm (O) =0.3$,
our observed 3 $\sigma$ upper limit on the water column density
is evidently consistent with the results of Williams et al.\ (and indeed
with any value of $f \rm (H_2O) \le 0.15$). Our results leave open the
possibility that both laboratory results are consistent, but that some water
measured by Vejby-Christensen et al.~(1997) is in a pre-dissociated state
(Adams 1997, private communication).

\section{Summary}

We have reported a $3\sigma$ upper limit of $9\times 10^{12}$ cm$^{-2}$
on the column density of water toward
HD 154368. We have constructed detailed chemical models which incorporate many
existing constraints on the physical conditions of the ambient medium.
Together with the known column density of OH along the line of sight, we
constrain the H$_2$O branching ratio for dissociative recombination of
H$_3$O$^+$ to be smaller than 30\%. Our results are in agreement with
the laboratory studies of Williams et al.\, which find an oxygen 
channel in the recombination 
of H$_3$O$^+$ and a small branching ratio for the production of water, 
but are mildly inconsistent with the laboratory results of Vejby-Christensen
et al.

The observed spectrum of HD 154368
also yielded high-resolution observations
of the Mg II doublet at 1239.9 \AA\ and 1240.4 \AA, allowing
the velocity structure of the dominant ionization state
of magnesium to be studied along the line-of-sight.  The
Mg II spectrum is consistent with GHRS observations at lower
spectral resolution that were obtained previously but
allow an additional velocity component to be identified.
Absorption by the Ge II line at 1237.059 \AA\ was also detected,
the first detection of interstellar germanium along the line-of-sight 
to HD 154368.

We acknowledge with gratitude the support of NASA grant NAGW-3147 from the
Long Term Space Astrophysics Research Program and of 
grant GO-06739.01-95A
from the HST Cycle 6 Program. 
In the final stages of this project
MS was partially supported by NASA through grant HF-01101.01-97A,
awarded by the Space Telescope Institute, which is
operated by the Association of Universities for Research in Astronomy,
Inc., for NASA under contract NAS 5-26555.
Most of the simulations presented in this work were performed on the Cray YMP
operated by the Netherlands Council for Supercomputer Facilities in Amsterdam.

\newpage
\begin{tabular}{c c c c c}
\multicolumn{5}{c}{Table 1}\\
\multicolumn{5}{c}{Velocities, $b$ values, Equivalent Widths,}\\
\multicolumn{5}{c}{and Column Densities of Mg II for HD 154368}\\
\tableline
\tableline
$V_{\rm helio}$ (km s$^{-1}$)&$b$ (km s$^{-1}$)&EW (m\AA)&EW (m\AA)
&log N (cm$^{-2}$)\\
&&($\lambda$ 1239.9)&($\lambda$ 1240.4)\\
\tableline
-36.2&1.2&1.9&1.3&13.7  \\
-26.7&1.3&1.8&1.2&13.6 \\
-16.9&2.7&19.6&13.3&15.0 \\
-1.1&2.9&39.3&26.9&16.2  \\
+10.3&1.5&3.6&2.4&14.0  \\
\tableline
\end{tabular}

\newpage

\centerline{\bf Figure Captions}
\vskip 0.1 true in 
{\hoffset -20pt
\parindent -20pt
Fig.\ 1a -- GHRS Echelle spectrum of HD 154386, showing the spectral region around
the expected position of the 1239.382 \AA, 1239.728 \AA\ and 1240.153 \AA\ 
water features.  Vertical lines show the line wavelengths for the
heliocentric velocity of the main Mg II component. The labels $1_{01}-2_{11}$,
$0_{00}-1_{10}$ and $1_{01}-1_{11}$ denote transitions of the
C\ $^1B_1$-X\ $^1A_1$ band of water.

Fig.\ 1b -- GHRS Echelle spectrum of HD 154386, showing the spectral region around
the expected position of the 1239.382 \AA\ water feature.  The vertical line
shows the line wavelength for the heliocentric
velocity of the main Mg II component.

Fig.\ 2 -- GHRS Echelle spectrum of the 
$\lambda$1239.925, 1240.395 Mg II doublet towards HD 154368. The dashed
curves indicate the fits.

Fig.\ 3 -- GHRS Echelle spectrum of the 
$\lambda$1237.059 Ge II line towards HD 154368. The vertical line shows the line
wavelength for the heliocentric velocity of the main Mg II component.

Fig.\ 4a --  Depth dependence of some major atomic and molecular species for
a model of the translucent cloud in front of HD 154368,
with $n_{\rm H}=325$ cm$^{-3}$, $I_{\rm UV}=3$, $A_{\rm V}=2.65$ mag
and the depletions of Snow et al.~(1996).

Fig.\ 4b. -- Variation of the gas kinetic temperature for the model in Fig 4a.

Fig. 5 -- Model grid for the column densities of OH and H$_2$O as functions
of $\zeta$ and $f({\rm H}_2{\rm O})$. The solid curves are for $f\rm (O) = 0$
and the dashed curve is for $f\rm (O) = 0.3$ (see text).} The cross shows the
observed OH column density (Black 1997, private communication) and our
3$\sigma$ upper limit on the H$_2$O column density.

\clearpage
\centerline{\psfig{figure=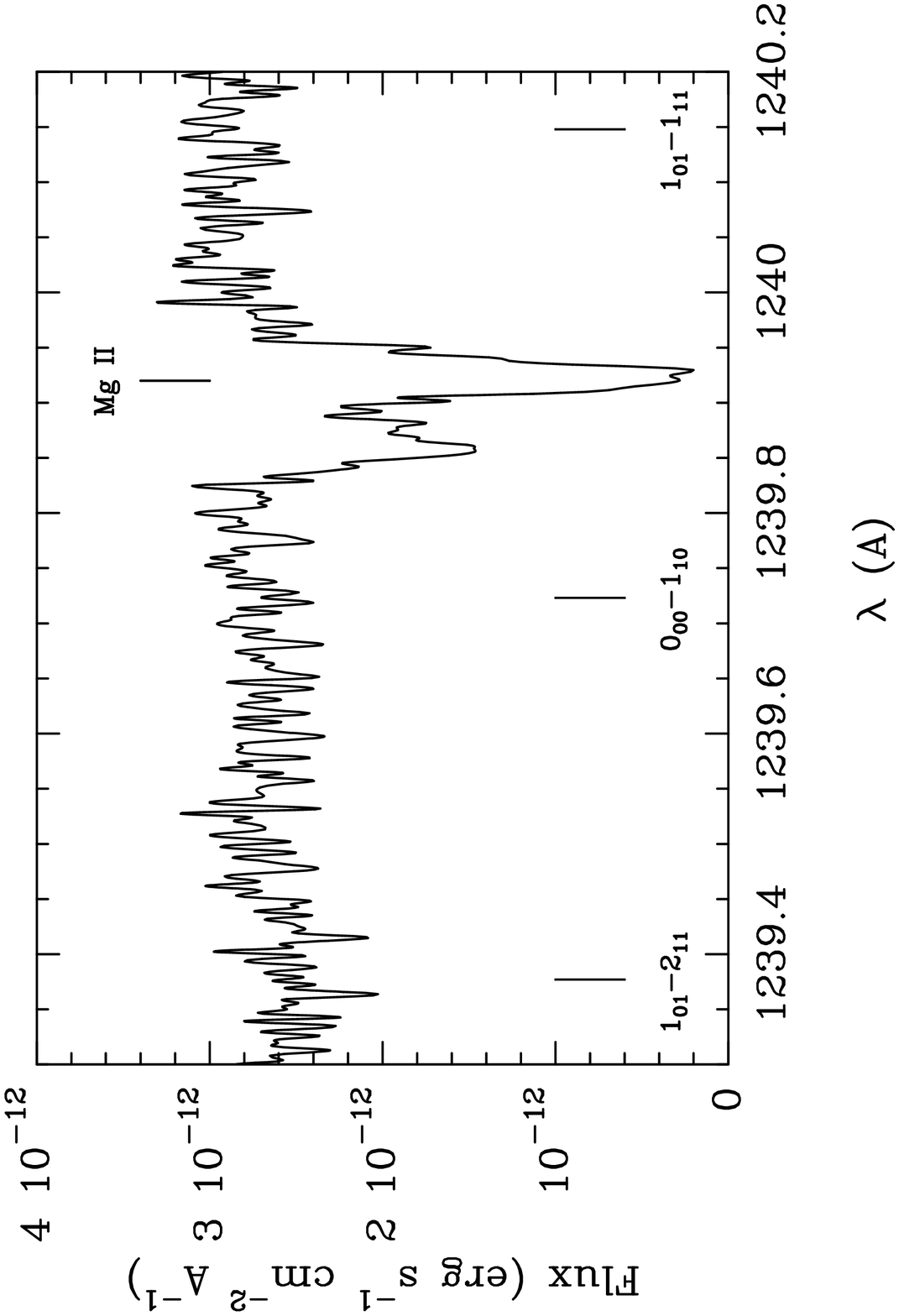,width=15.0truecm}}

\clearpage
\centerline{\psfig{figure=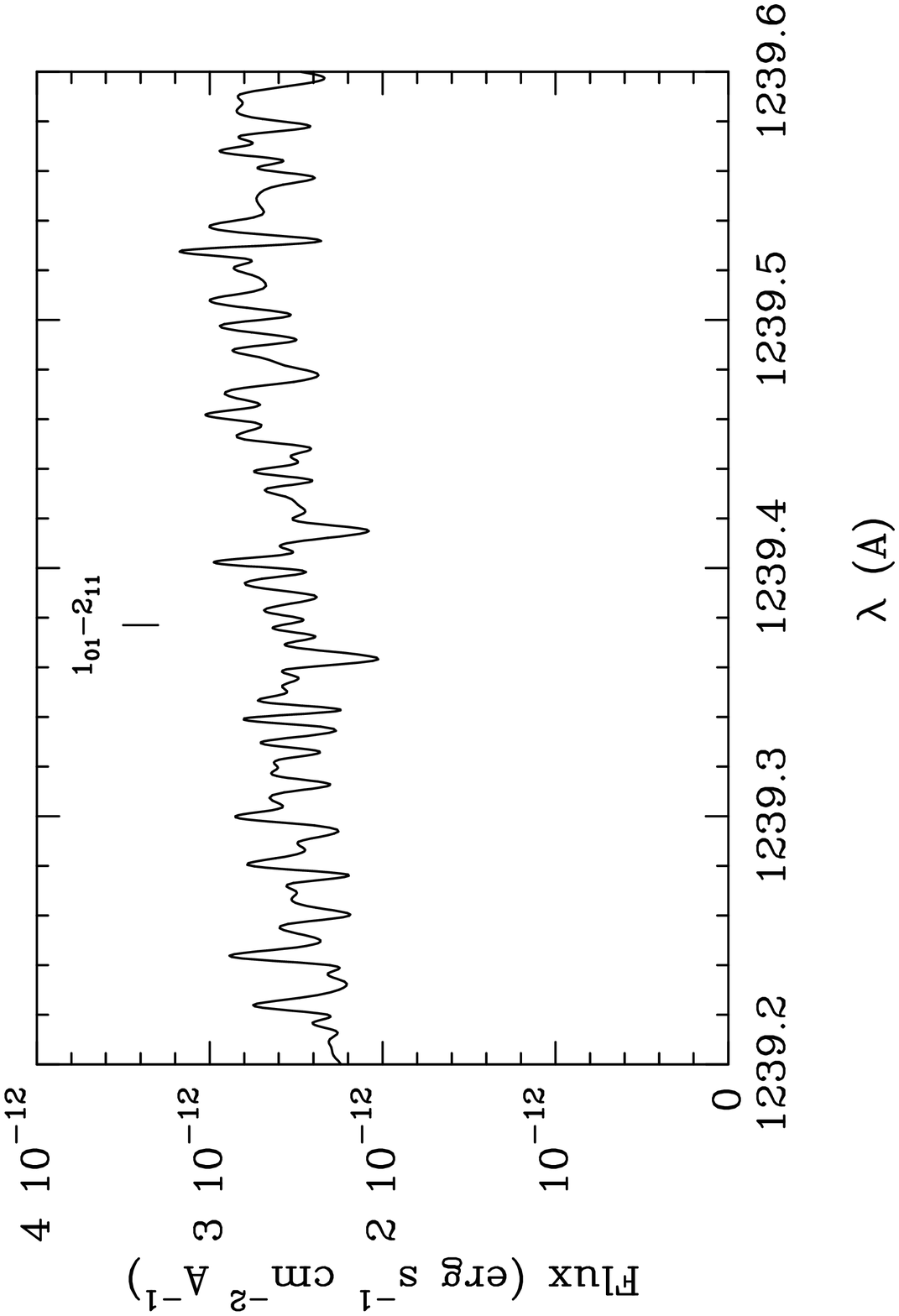,width=15.0truecm}}

\clearpage
\centerline{\psfig{figure=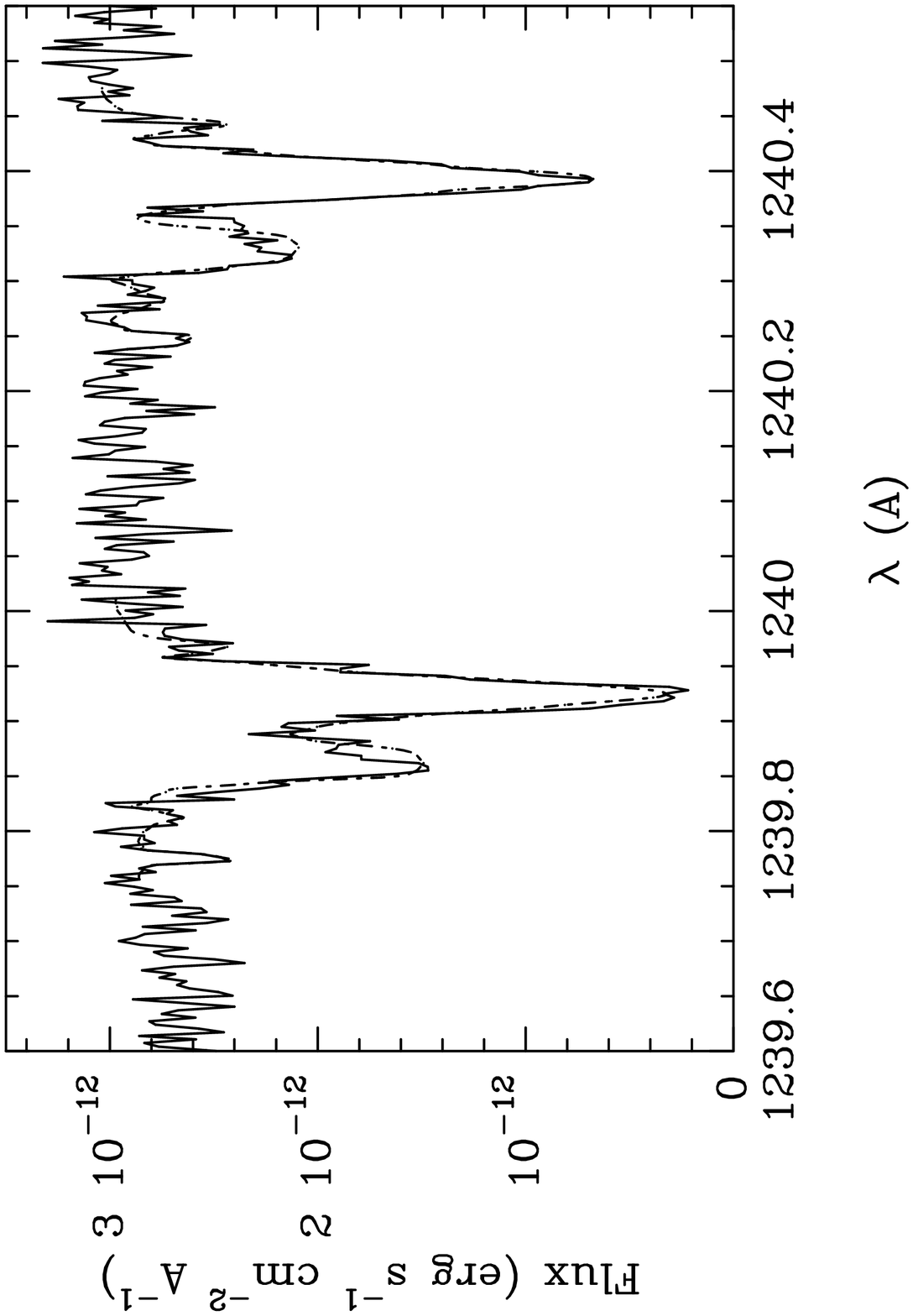,width=15.0truecm}}

\clearpage
\centerline{\psfig{figure=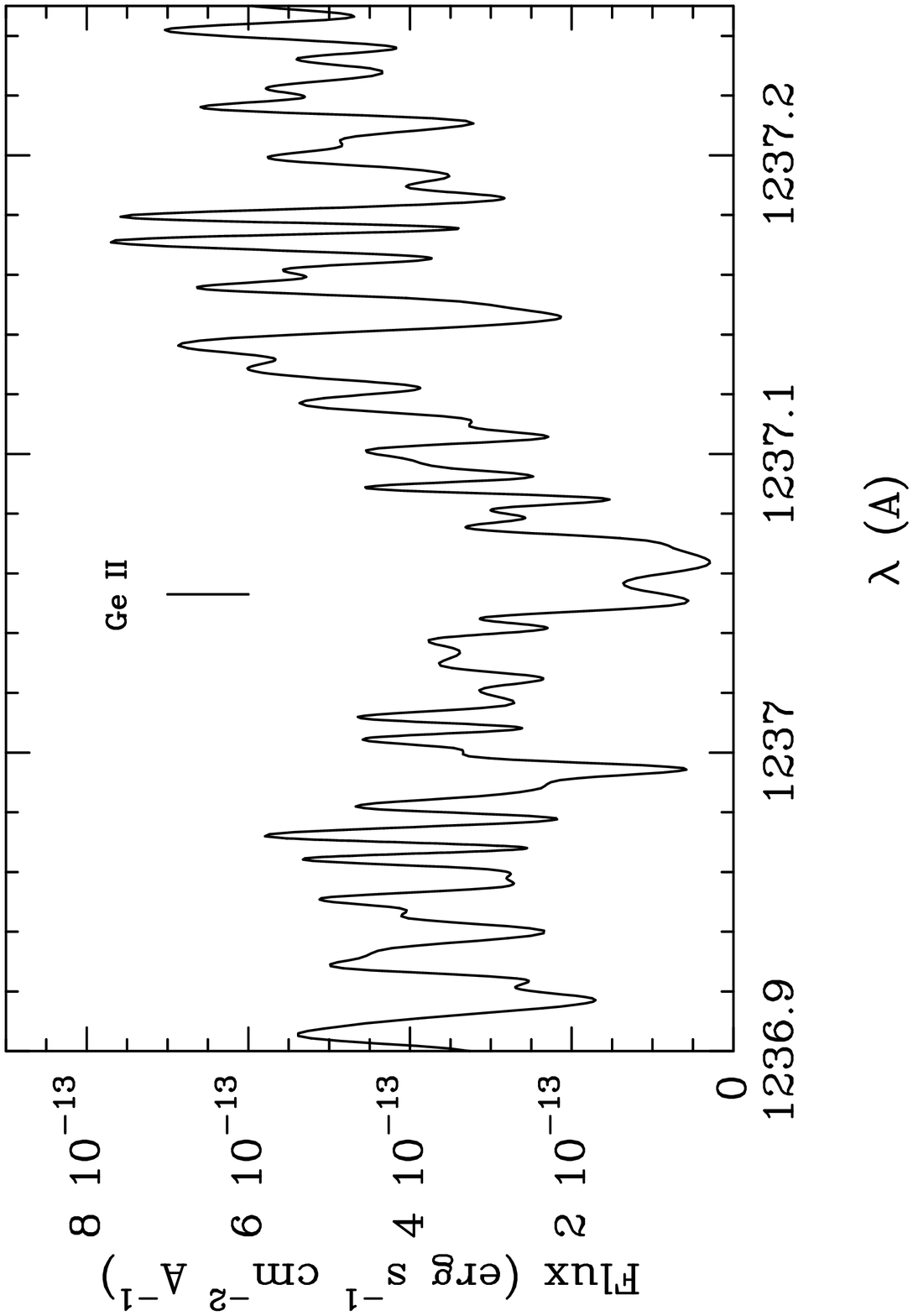,width=15.0truecm}}

\clearpage
\centerline{\psfig{figure=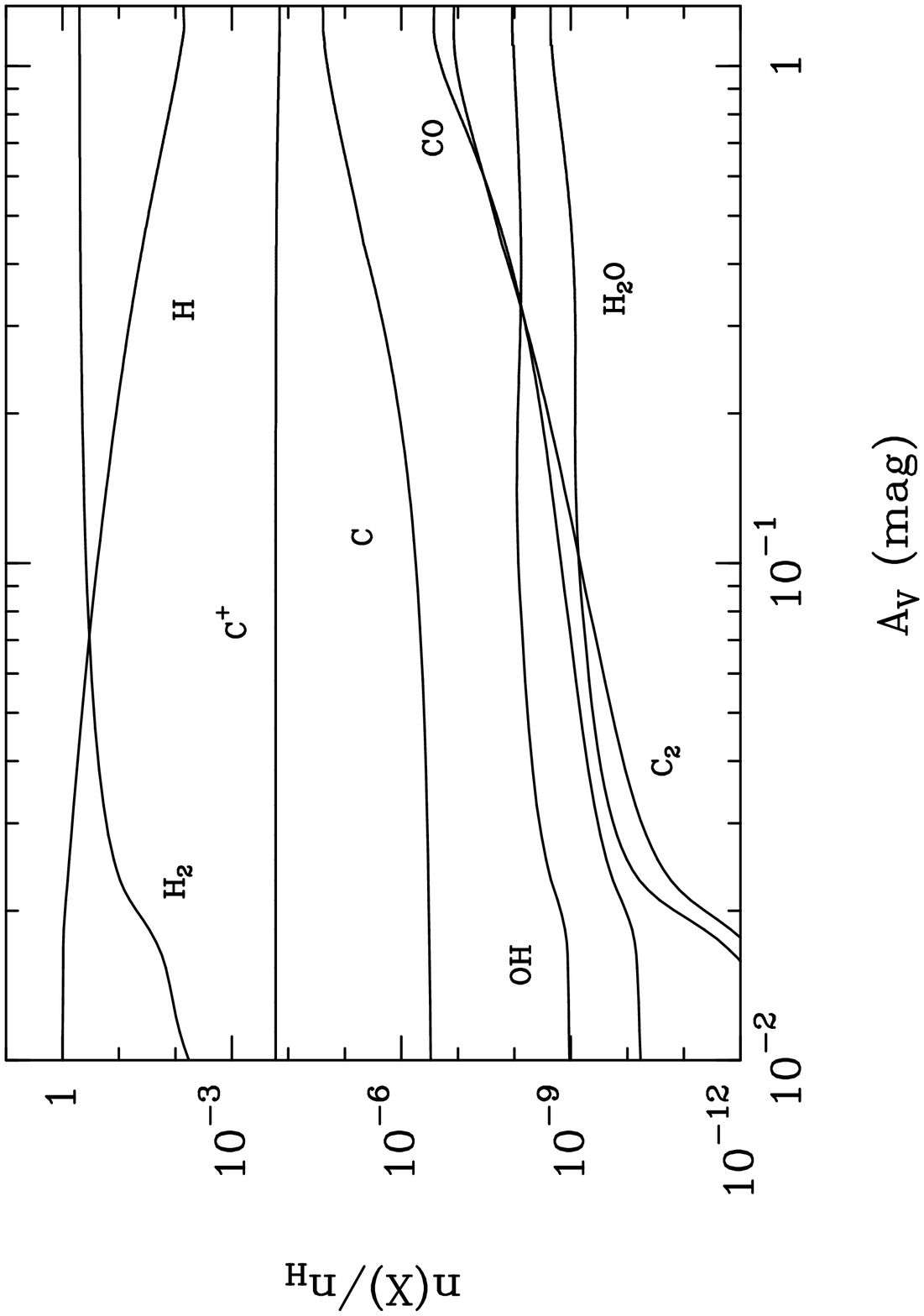,width=15.0truecm}}

\clearpage
\centerline{\psfig{figure=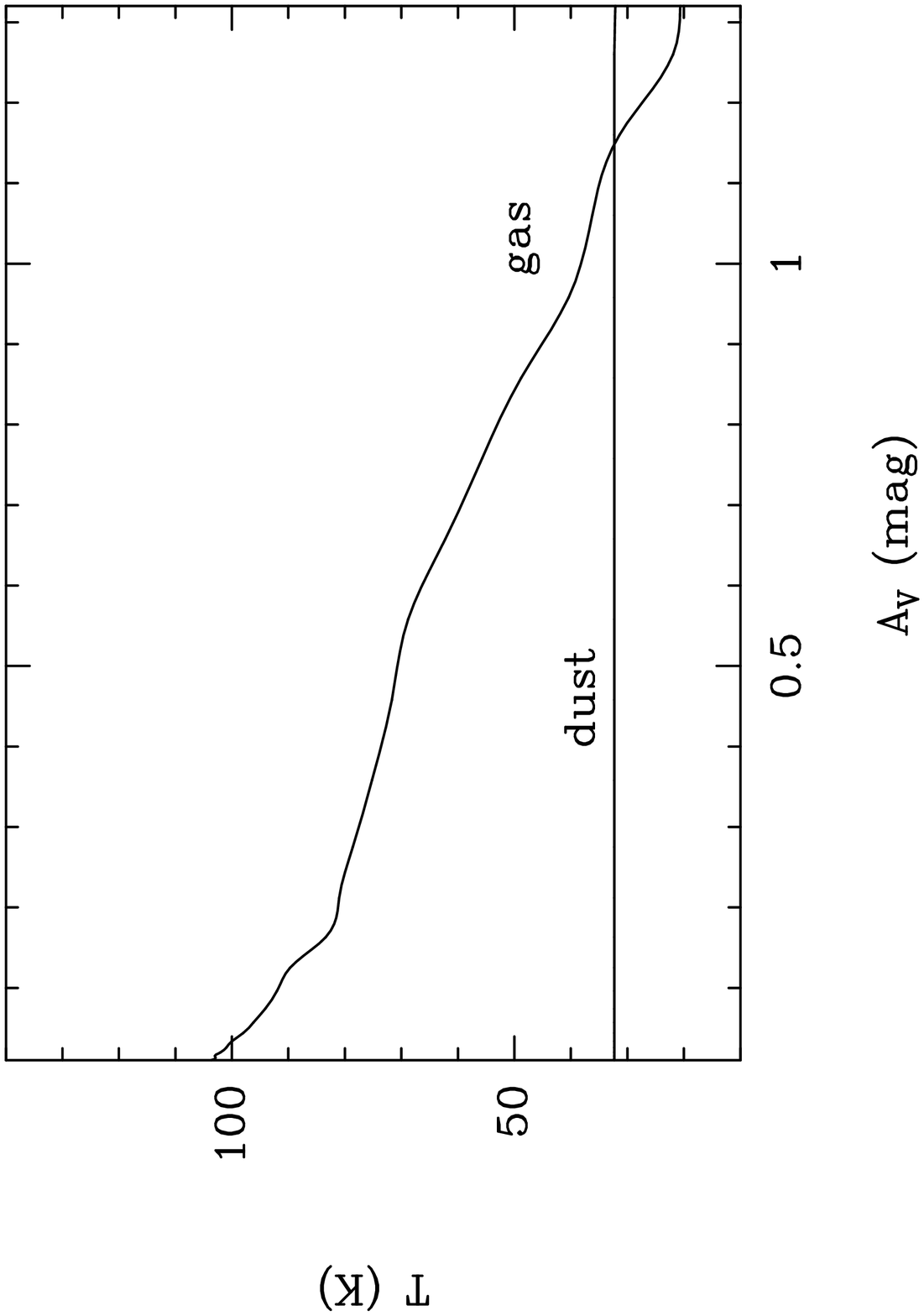,width=15.0truecm}}

\clearpage
\centerline{\psfig{figure=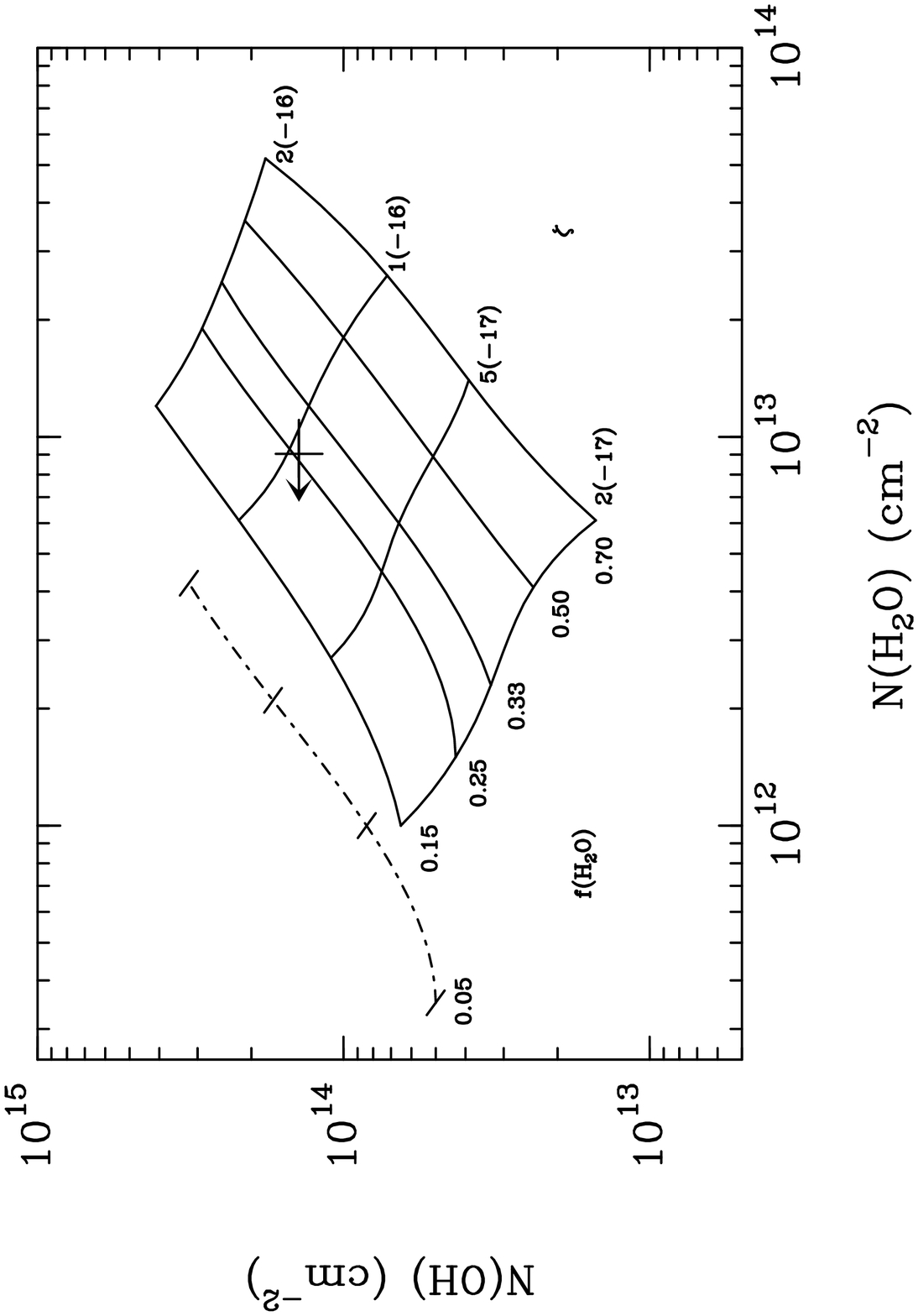,width=15.0truecm}}

\end{document}